\documentclass{vldb}
\pdfoutput=1
\AtBeginDocument{%
  \providecommand\BibTeX{{%
    \normalfont B\kern-0.5em{\scshape i\kern-0.25em b}\kern-0.8em\TeX}}}

\usepackage{cite}
\usepackage{balance}
\usepackage{blindtext}
\usepackage{comment}
\usepackage{amsmath}
\usepackage[inline]{enumitem}
\usepackage{array}
\usepackage{graphicx}
\usepackage{adjustbox}
\usepackage{subcaption}
\usepackage{tabularx}
\usepackage{ragged2e}
\usepackage{multirow}
\usepackage{booktabs}
\usepackage{caption}
\usepackage{float}
\usepackage{url}
\usepackage{xspace}

\usepackage[ruled,norelsize]{algorithm2e}

\vldbVolume{}
\vldbNumber{}
\vldbYear{}

\newcommand{\name}{\namex\xspace}

\newcommand{\Paragraph}[1]{~\vspace*{-0.8\baselineskip}\\{\bf #1}}

\newtheorem{problem}{Problem}

\newcommand{\bm}[1]{\mathbf{#1}}

\DeclareMathOperator{\Sm}{\mathcal{S}}
\DeclareMathOperator{\Cm}{\mathcal{C}}

\newcommand\blfootnote[1]{%
  \begingroup
  \renewcommand\thefootnote{}\footnote{#1 \vspace{-7mm}}%
  \addtocounter{footnote}{-1}%
  \endgroup
}

\newenvironment{pkl}{%
\begin{itemize}%
\setlength\itemsep{-0.5\parskip}%
\setlength\parsep{0in}%
}{%
\end{itemize}}

\vldbTitle{ Database Workload Characterization with Query Plan Encoders }
\vldbAuthors{Debjyoti Paul, Jie Cao, Feifei Li, Vivek Srikumar }
\vldbDOI{https://doi.org/TBD}

\begin{document}


\title{ Database Workload Characterization with Query Plan Encoders}




%
%
%
%

\numberofauthors{1} 

\author{
%
%
\alignauthor
Debjyoti Paul$^\dagger$, Jie Cao$^\dagger$, Feifei Li, Vivek Srikumar\\
       \affaddr{School of Computing, University of Utah}\\
       \affaddr{\{deb, jcao, lifeifei, svivek\}@cs.utah.edu}
}

\maketitle
\begin{abstract}
	Smart databases are adopting artificial intelligence (AI) technologies to
	achieve {\em instance optimality}, and in the future, databases will
	come with prepackaged AI models within their core components. The
	reason is that every database runs on different workloads, demands
	specific resources, and settings to achieve optimal
	performance. It prompts the necessity to understand workloads running
	in the system along with their features comprehensively, which we dub as workload characterization.

	To address this workload characterization problem, we propose our query plan
	encoders that learn essential features and their correlations from
	query plans. Our pretrained encoders captures the {\em structural} and
	the {\em computational performance} of queries independently. We
	show that our pretrained encoders are adaptable to workloads that
	expedites the transfer learning process.  We performed independent
	assessments of structural encoder and performance encoders with
	multiple downstream tasks. For the overall evaluation of our query plan
	encoders, we architect two downstream tasks (i) query latency prediction
	and (ii) query classification. These tasks show the
	importance of feature-based workload characterization. We also
	performed extensive experiments on individual encoders to verify the
	effectiveness of representation learning, and domain adaptability.
\end{abstract}


\blfootnote{$^\dagger$ These authors contributed equally.}
\vspace{-3mm}
\section{Introduction}
\label{sec:introduction}
Database Management Systems (DBMS) are general-purpose systems that aim to
provide solutions to many applications as possible. Database designers expose
many configuration settings to facilitate end-users in managing complex
workloads efficiently. However, there is no single configuration that works for
all workloads and finding the optimal configuration setting is very dependent
on the workload characteristics.

In the usual process, DBAs first need to learn about the database queries that
frequently run on their database system and then dig deeper to characterize
these queries. It requires an in-depth knowledge and robust understanding of
the queries and its execution features.  It is a challenging as well as
laborious task for DBAs to comprehend execution features of queries and its
relations with configuration knobs. Furthermore, the large number of possible
DBMS configurations settings makes it a daunting task for DBAs.  Currently,
advanced DBAs applies simple data mining techniques and hand-tweaked feature
engineering to understand the nature of workload, but this requires
domain expertise, which is rare.

Nowadays, many small to medium businesses (SMBs) manage their databases with
cloud services. Cloud database service providers can now obtain and analyze
large amounts of anonymized workload data. Managing database resources
efficiently is indispensable for providing quality services. Each database
instance runs a different workload.  Applying data science can help in
identifying workloads with similar characteristics, and then it can be used in
downstream tasks e.g., query optimization, configuration recommendation and
index recommendation. Essentially, it raises a requirement of database workload
characterization, i.e.,  ability to describe distinctive nature and features of
queries in a workload.

Previous work \cite{thummala2010ituned} shows with TPC-H benchmarks how each
database query behaves differently with changes in database configuration
settings.  For example, query Q18 and query Q7 in TPC-H benchmark responds to
knob changes \texttt{ shared\_buffers} vs. \texttt{effective\_cache\_size} very
differently w.r.t. {\em query latency}. Each query possesses distinct features,
and the demands for computational resources are also different. It suggests
that each query needs to be treated uniquely and based on their characteristic.
Recent research works \cite{ding2019ai,ding2018plan,li2019qtune} leverages
query plans as the {\em feature description of queries} and use it for tasks
like index recommendation \cite{ding2019ai,ding2018plan} and configuration knob
tuning \cite{li2019qtune}.

In natural language domain, a word is a structural and functional unit of a
meaningful sentence. Similarly, in the database domain, if a query is the
structural unit, then a database query plan is the functional unit of a
workload. With the advancement in distributed representation of words, the
downstream tasks like sentence similarity, question answering, and textual
entailment have improved dramatically \cite{mikolov2013distributed,
devlin2018bert, yang2019xlnet}. In a similar way, we foresee that downstream
tasks like workload similarity, index recommendation, and database
configuration recommendation can benefit from the study of workload
characterization.

We propose a scalable data-driven artificial intelligence (AI) approach for
workload characterization with distributed representation of query plans.  One
of the benefits of AI deep learning models is automatic feature engineering and
auto-correlation among features.  It is an non-trivial arduous task and possess
many challenges in achieving the aim of workload characterization. Some of the
challenges that makes it very different from other entity represention learning
are {\em Query Independence, Diverse Query Structure, Modeling Computational
Complexity,} and {\em Data Dependence}. We present a constructive detail on
each of aforementioned challenges in \S\ref{sec:challenges}.

\Paragraph{Our Approach.} In our work, we first propose a query plan
distributed representation model that captures the inherent characteristics
such as {\em structure, computational demand}, and {\em feature manifests}
embedded within a query plan structure. Hence, we created two parts for query
plan representation, {\em (i)  Structure Representation, (ii) Computational
Performance Representation}. The two representations, either separately or
collectively, can be used in downstream tasks to understand a query
comprehensively. As an example, we demonstrate an approach to perform query
latency prediction with the help of query representations. It can help in
offline profiling of workloads and aid in tuning database settings. We believe
that instance optimality of a database can only be achieved with the in-depth
understanding of queries running in a system, and suggests the introduction of
workload characterization component for it.

In our choice of design for distributed representation, we can either use a
{\em fixed-embedding} or a {\em pretrained encoder} approach. Fixed embedding
is useful where the set of elements is complete, and after model training, we
get a fixed representation for all the elements in the set. This approach is
instrumental in domain like graph embedding. On the other hand, pretrained
encoder is a learned model that can output embedding on receiving the input by
featurizing the attributes from the input along with learned weights from
previous observations. We follow the pretrained encoder approach for
adaptibility and transfer of knowledge.

Furthermore, we follow a bidirectional encoder strategy with both {\em
feature-based} and {\em finetuning-based} approach inspired by the language
models \cite{devlin2018bert}.  In this approach, the embedding obtained from
the pretrained encoder is trained to learn features, and then the feature
embedding output can be fed to multiple task-specific models.
The approach aims to alleviates the requirement of task-
specific representation and facilitates reuse of already learned features from
the model to multiple domain-specific task. A pre-trained plan representation
model also simplifies the transfer learning process when trained on a large
dataset and fine-tuned for specific data and problem set.


We summarize the contribution of this paper.
\vspace{-3mm}
\begin{pkl}
\item We propose plan encoders for distributed representation of query plans.
        The general feature-based encoders capture inherent characteristics of
        query plans.

\item We capture two aspects of the query plans independently with two classes
        of encoders. The {\em structure }, and the {\em computation} of query
        plans.

\item The {\em structure} encoder is inspired by the natural language model,
        representing a tree-structure of heterogeneous operators in a latent
        multidimensional space. Thereafter, we evaluate our structure encoder
        model with similar query classification and regression tasks on
        multiple datasets.

\item Our {\em computational} encoder is a collection of encoder instances.
        Each encoder corresponds to a database operator such as scan, join,
        sort, aggregate, etc., optimizing for multiple metrics to capture the
        computational features. The encoder uses statistical information and
        data distribution of the underlying relational data along with the
        explicitly specified plan features and database configurations.

\item The train both our encoders with a large dataset of diverse query plans
        and benchmarks for pretraining. We then introduce a finetuning-based
        approach that can quickly adapt to new data distribution with limited
        data resources. It is essential for increment learning and fast domain
        adaptation with new workloads.

\item To show the overall effectiveness of our encoders, we performed query
        latency prediction and query
        classification tasks. In query latency prediction, given a query plan and a database configuration setting, the downstream model predicts the query latency using our plan encoders. In query classification task, we use our plan encoders to classify closely related queries.
\end{pkl}
\vspace{-2mm}

The rest of the paper is organized as follows. \S\ref{sec:preliminaries}
provides background and challenges we face while performing query plan
representation, respectively. In \S\ref{sec:model}, we present our
structure encoder and performance encoder, followed by downstream tasks using
plan encoders in \S\ref{sec:downstream_tasks}. We present experiments and
results of our downstream tasks with plan encoders in
\S\ref{sec:experiments}, and analysis of individual encoders in \S\ref{sec:analysis}. We present a brief section on related works in \S\ref{sec:related}, followed by
conclusion in \S\ref{conclusion}.


\vspace{-2mm}
\section{Preliminaries}
\label{sec:preliminaries}
Recently we are noticing a trend of utilizing the power of Artificial
Intelligence (AI) in buffer resource tuning, indexing, and query optimizer
\cite{kraska2019sagedb, tan2019ibtune, marcus2019plan}. In the near future, we
expect database systems packaged with pretrained AI models, and dedicated cloud
servers with embedded AI accelerators to facilitate the processing. Our
proposed workload characterization with distributed representation of query
plans can empower database core components to operate efficiently with in-depth
insights on workloads.

\begin{figure*}[!t]
    \vspace{-3mm}
    \begin{minipage}[t]{0.33\textwidth}
    \centering
    \includegraphics[width=\textwidth, height=7.2cm]{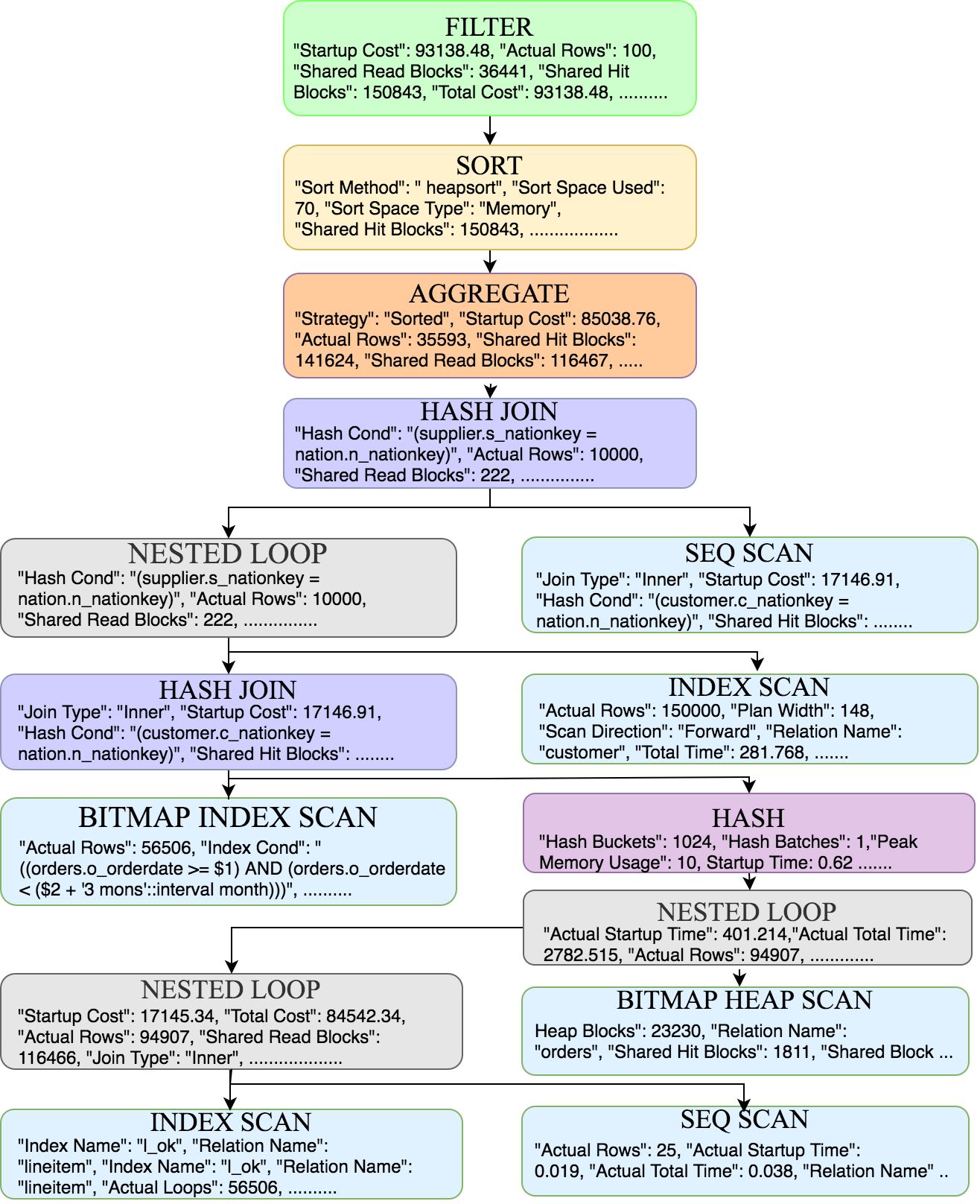}
    \captionsetup{justification=centering}
    \vspace{-4mm}
    \captionof{figure}{\small A query execution plan from TPC-H\cite{bench:tpch}.}
    \label{fig:plan_01}
    \end{minipage}
    \begin{minipage}{0.67\textwidth}
        \vspace{-6.1cm}
      {\small
      \begin{adjustbox}{width=0.95\textwidth,center}
      \begin{tabular}{lp{7cm}} \toprule
      \textbf{Operator} & \textbf{Plan Properties or Features} \\ \midrule
        \multirow{5}{*}{All}  & Actual Loops ,  Actual Rows ,  Local Dirtied Blocks ,  Local Hit Blocks ,  Local Read Blocks ,  Local Written Blocks ,  Plan Rows ,  Plan Width ,  Shared Dirtied Blocks ,  Shared Hit Blocks ,  Shared Read Blocks ,  Shared Written Blocks ,  Temp Read Blocks ,  Temp Written Blocks ,  Parent Relationship ,  Plan Buffers \\ \hline
        \multirow{3}{*}{Scan} & Relation Name, Scan Direction, Index Name, Index Condition, Scan Condition, Filter, Rows Removed, Heap Blocks, Parallel, Recheck Condition	 \\ \hline
        \multirow{3}{*}{Join} & Join Type, Inner Unique, Merge Condition, Hash Condition, Rows Removed by Join Filter, Parent Relationship, Hash Algorithm, Hash Algo, Hash Buckets, Hash Batches, Peak Memory \\ \hline
        \multirow{2}{*}{Sort} & Sort Type, Sort Method, Sort Space, Sort Key, Sort Space Type, Sort Space Used, Peak Memory \\ \hline
        \multirow{2}{*}{Aggregate} & Strategy, Hash Algo, Hash Buckets, Hash Batches, Parallel Aware, Partial Mode, Peak Memory\\ \bottomrule
      \end{tabular}
      \end{adjustbox}
      \captionsetup{justification=centering}
      \vspace{-2mm}
      \captionof{table}{\small The properties from query execution plan that are common to all the operators and a few specific to major operators like Scan, Join, Sort and Aggregate.}
      \label{table:plan_prop}
        }
    \end{minipage}
    \vspace{-5mm}
\end{figure*}

\vspace{-1mm}
\subsection{Workload, Query and Query Plan.}
We define a database workload as $W =\{(q_1,\theta_1), (q_2,\theta_2),$
$\ldots,(q_{n},\theta_{n})\}$, where $q_i$ is the database query, and
$\theta_i$ is a normalized weight of importance of $q_i$ in workload $W$ such
that $\sum_{i=1}^n \theta_i = 1$. The weight $\theta_i$ can be as simple as the
frequency of appearance of $q_i$ in $W$  or can be arbitrarily decided by the
DBA. Generally, database users mostly runs a set of predefined template queries
with seldom ad-hoc queries on databases. A data-driven smart database should
collect statistics of each query and use it to determine the importance of a
query.

For each query $q_i$, one can obtain the corresponding query plan $p_i$ from
the database system. Also, to note that a query with a similar template can
generate a different query execution plan or {\em query-plan} based on the
meta-information of a table in a database. Let us say,  $q_i$ generates two
query plan $p_k$ and query plan $p_l$ on different instances. In our approach,
it is safe to assume and treat both the queries different from the functional
point of view. Hence, there can be a one-to-many mapping from queries to
query-plans.

Alternatively, we can now define workload as $$W
=\{(p_1,\theta_1),(p_2,\theta_2),\ldots,(p_{m},\theta_{m})\},$$ where $p_i$ is
the database query-plan, and $\theta_i$ is a normalized weight of importance of
$p_i$ in workload $W$ such that $\sum_{i=1}^n \theta_i = 1$. For readability,
we will refer a query-plan as a plan in the paper from now.

A plan is a tree structure with heterogeneous functional operator nodes like
{\em Seq Scan, Index Scan, Bitmap Heap Scan, Nested Loop, Hash Join, Aggregate,
Sort, Filter} etc. Each operator node contains a set of execution properties.
We present an example of a plan in Figure \ref{fig:plan_01} of query Q5 from
the TPC-H benchmark with operator types.  All operators have a set of common
properties, and in addition, a few contain specific properties based on their
functions. These operator properties carry valuable information about their
execution.  Based on the functions of each operator, we grouped all operators
into five exclusive groups, i.e., {\em Scan, Join, Aggregate, Join} and {\em
Others}. In Table \ref{table:plan_prop}, we lay out the properties common to
all groups as 'All' and the properties exclusive to Scan, Join, Sort, and
Aggregate operators. These operator properties are used for computational
performance representation of the plan. Please note that we do not use
properties like {\sf Total Cost, Actual Total Time, Actual Startup Cost}
because we use them as labels in our prediction tasks. We describe it in
\S\ref{sec:computational}.

For any plan $p_i$ as input to our {\em Structural Encoder} and {\em
Computational Encoder}, the models outputs the  structural embedding $\Sm(p_i)$
and the computational performance embedding $\Cm(p_i)$, respectively. These
embeddings are used by downstream models for different fine-tuning tasks.

\vspace{-1mm}
\subsection{Deep Neural Networks (DNNs)}
DNNs are widely used computational frameworks for many AI applications. DNNs
are layers of neuron thoughtfully structured that performs a weighted sum
computation of the input values at each neuron. A structure of DNNs or {\em
model} is also an instance of a machine learning algorithm that learns patterns
from data by inferencing and then readjusting weights to minimize error.  DNNs
are very efficient in reducing high dimensional data into low dimensional code
or features \cite{hinton2006reducing}. DNN hardly requires feature engineering
and can learn complex relations among multiple features. In our paper, we are
specifically interested in the entity representation learning capability of
DNNs.  Moreover, we focus our attention on Autoencoder (an Encoder-Decoder
approach) for learning the {\em structural} representation model. A particular
kind of Autoencoder called Denoising Autoencoder can capture robust generalized
features from original data \cite{vincent2008extracting}. We applied an
advanced feature-based encoding and learning technique inspired by natural
language models.  Recent applications of encoder architectures on language
models are very successful in capturing structural and statistical properties
\cite{devlin2018bert, yang2019xlnet}. Query plans are structurally complex, and
properties of plan operators are implicitly correlated. Hence, we adapted the
autoencoder approach in our representation models.  For the {\em computational
performance} representation, we used a supervised learning approach learning
features contributing to multiple metrics for operators.

\vspace{-1mm}
\subsection{Challenges and Mitigation Strategies}
\label{sec:challenges}
Traditional machine learning approaches encode entities into a fixed-length
features before feeding them into any model for prediction tasks.  We provide a
consolidated set of challenges we face while performing workload
characterization with plan encoders because of heterogeneous nature, diverse
shape, and varying depth of plans.

\vspace{-2mm}
\begin{pkl}
\item {\em Query Independence:}
Each query is unique and independent. Even if the queries are from the same
        benchmark or workload, they are seldom similar in structural and
        computational complexity. Unlike other entity embedding where
        contextual appearances of entities play a pivotal importance (such as
        word embedding), in workload contextual or temporal appearance of
        queries are not related.

\item{\em Diverse Query Structure:} The structure of query plans is represented
        as a tree of functional operator nodes, e.g., {\em scan, join, sort}
        etc. It is a non-trivial task to represent a tree structure containing
        attribute features at every node.

\item{\em Modeling Computational Complexity:}  Each query has a specific demand
        for computational resources based on their functional operations.
        Moreover, the resource demand of each functional operator is different.
        An open question arises whether to implement an operator-level model or
        a single primary model for encoding.

\item{\em Data Dependence:} In databases, the generation of query plans from a query
        depends on many factors, such as index availability, statistical
        information on data. A complete query plan can only capture basic
        information about underlying data. It arises a question whether it is
        enough or we need to incorporate more information.

\item{\em Encoding Multiple Properties:}
Database plans contain interrelated properties and information that give hints
        about query performance and their execution metrics such as latency and
        throughput. It is a challenge to unify and discover complex correlation
        among the properties and features explicitly obtained from plans.

\item{\em Domain Adaptation:}
The encoder models trained on a set of workloads are likely to encounter a
        different unseen workload in the prediction phase. It is a challenge to
        adapt to a new workload setting quickly (with less training data) using
        the prior pretrained weights of the models.
\end{pkl}
\vspace{-4mm}

We adopted specific strategies in our approach to address the above challenges.
We purposefully design a feature-based query plan encoder for learning the
individual characteristics from different query plans. For modeling the
performance complexity, we incorporate meta-information (e.g., data
distribution, selectivity, cardinality) of database tables and attributes used
in queries providing a detailed picture of the data access pattern.

In our approach, it is a not trivial attempt to incorporate all the relevant
meta-information and capture relevant features in our plan performance
representation. Still, it is reasonable to assume that if we can incorporate
all the required information to the encoders, then we might be able to learn
the influencing factors contributes to evaluation metrics of query plans. After
all, the query optimizers are universally designed logical components that
generate query plans.  The encoders producing distributed representation of
query plans can facilitate many downstream tasks and enhance the performance of
core components. It encourages us to keep the encoder as general as possible and capture the
correlation among properties well enough in the query plan representation.
With a data-driven approach, we aim to create a pretrained encoder model that
learns from large and diverse datasets to learn plan features.  In ideal scenario, we want pretrained encoders to quickly adapt new domain
with less dataset, expediting domain transfer.


\vspace{-4mm}
\section{Query Plan Representation}
\label{sec:model}
In this section, we present our {\em Structure Encoder} and {\em
  Computational Performance Encoder} for plans.  Each node in the tree
is an instance of a functional operator with multiple properties, and
nodes are ordered and connected via unlabeled edges depicting the
dependence relation. For structural representation, we mainly study
the operator type of each node and leaving the performance-related
properties for {\em computational performance representation} in
\S\ref{sec:computational}.  When sketching our encoders, we
realize keeping the {\em structure}, and {\em computational
  performance} representation independent increases the modularity in
design, which also enables downstream tasks to choose and weigh each
representations independently in their model. It also helps us in
evaluating the structure and performance encoders separately.

For both Structure Encoder and Computational Encoder, we hope that our
pretrained model can be easily adapted to new applications. Hence, we
study both of them on a two-stage framework: pretraining and
finetuning.  In this section, we mainly introduce the pretraining
tasks and model architectures for them. Then we outline our
finetuning evaluation in \S\ref{ssec:finetuning}.

\begin{figure*}[t]
    \centering
    \includegraphics[width=0.95\textwidth]{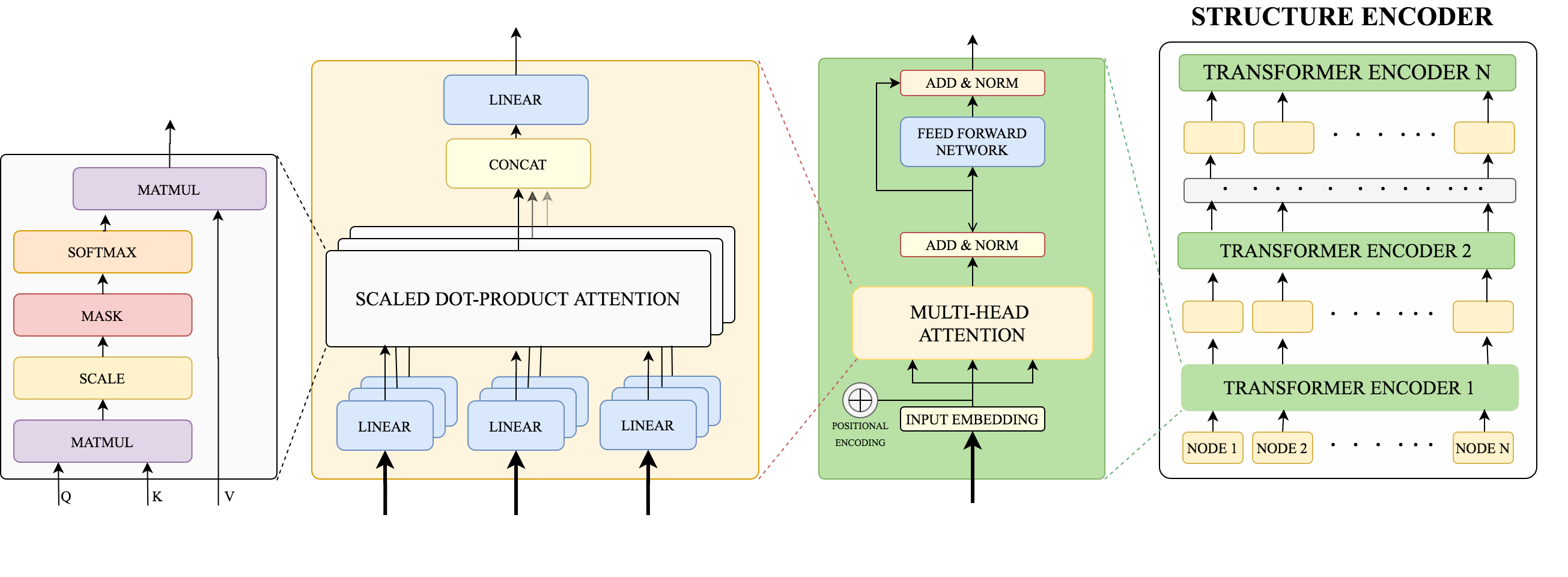}
    \captionsetup{justification=centering}
    \vspace{-7mm}
    \captionof{figure}{Structure Representation Model.}
    \label{fig:structure}
    \vspace{-5mm}
\end{figure*}

\vspace{-1mm}
\subsection{Structure Encoder}
\label{sec:structure}
Earlier, we have mentioned about the heterogeneity of operators in plans. We
now try to give a clear picture of the diverse types of operators plans can
possess. Same functional operators can use different strategies to fulfill
their operations. There are multiple types of Scan operators like Sequential
Scan, Index Scan, Bitmap Heap Scan, etc. Again, the same strategy often used in
multiple functional operators, like, Hash Join and Hash Aggregate uses Hash
strategy. We organized each type of operator into three sub-level type as
taxonomy of operators. The top-level {\em Level 1}, mostly suggest the
functional property such as Sort, Insert, Union, Scan, Join, etc. {\em Level 2}
and {\em Level 3} are grouped based on mutually exclusive strategy types such
as Hash, Index, Heap, etc. Table \ref{table:type_dict} shows all three levels
of operator sub-type for defining a real operator. We define all operator with
three sub-type as $\langle$Level 1$\rangle$-$\langle$Level
2$\rangle$-$\langle$Level 3$\rangle$. For example, operator {\em Bitmap Heap
Scan} and {\em Left Merge Join} is represented as Scan-Heap-Bitmap and
Join-Merge-Left, respectively. All those operaters types forms the tree
structure as shown in Figure \ref{fig:plan_01}, we need to find a way to encode
the tree. Notice that workload analysis based on similar query plans can help
DBAs in optimal utilization of database resources, e.g., buffers and
configuration, by utilizing historical experiences from other databases.
Furthermore, encoders enables clustering of similar-featured queries learned
from a large set of queries without actually sharing any private/sensitive
query information. Inspired by this goal, we propose a plan-pair similarity
regression task to guide the structural representation learning.

\vspace{-1mm}
\subsubsection{Plan-pair Similarity Regression}
\label{sssec:structure_pretraining}
To measure the similarity between two plans, we adopt a widely used graph
similarity metric in natual language representation domain: {\em
Smatch}~\cite{cai2013smatch}. It caculates the degree of overlap between two
graph structures, which is defined as the maximum F1-score obtainable via a
one-to-one matching of each node in two graph. Hence it is value from 0 to 1, 0
means totally different, while 1 means exact the same. In this task, we treat
the optimal Smatch score as the similiary of two plans. The Smatch score
between two tree-structure plans can be computed by graph macthing optimization
algorithm, such as Interger Linear Programming~(ILP) or Hill-climbing methods.
After we get the {\em Smatch} scores $s_{ij}$ of each plan-pair $<p_{i},
p_{j}>$, this can easily form a large dataset with {\em Smatch} score as the
similarity supervision. Based on thss large dataset, we propose that
pretraining our structure encoder with a task of predicting the {\em Smatch}
score of each plan pair. Results on the downstream similarity tasks or other
applications also shows that the structure encoder pretrained from this task
can be easily finetuned on new application or new domain.

\begin{table}[t]
  \begin{center}
    {\small
      \begin{tabular}{lp{6.8cm}} \toprule
        \bf{Level}               & \bf{Operator Sub-types}                                                                                                                                                                                                                                                      \\ \hline
        \multirow{5}{*}{Level 1} & Aggregate, Append, Count, Delete, Enum, Gather, Aggregate (Group, GroupAggregate), Hash, Insert, Intersect, Join (Nested Loop), Limit, LockRows, Loop, ModifyTable, Network, Result, Scan, Sequence, Set(SetOp), Sort, Union, Unique, Update, Window, WindowAgg, Materialize \\ \hline
        \multirow{3}{*}{Level 2} & And, CTE, Except, Exists, Foreign, Hash, Heap, Index, IndexOnly, LoopHash, Merge, Or, Query, Quick, Seq, SetOp, Subquery, Table, WorkTable                                                                                                                                   \\ \hline
        \multirow{1}{*}{Level 3} & Anti, Bitmap, Full, Left, Parallel, Partial, Partition, Right, Semi, XN                                                                                                                                                                                                      \\ \bottomrule
      \end{tabular}
    }
  \end{center}
  \vspace{-5mm}
  \caption{\label{table:type_dict} The taxonomy of operator types for every node}
  \vspace{-4mm}
\end{table}

\subsubsection{Model Architecture}
\label{sssec:structure_model}
In this paper, by linearizing the tree structure into a sequence of nodes, we
can transform the tree-encoding problem into a sequence encoding problem, which
has been well studied in many areas, such as natural language processing, time
serials analysis.

\Paragraph{Tree Traversal} When linearize the tree into node sequence. We use
the root first traversaling, but adding hierarchical brackets for each
non-terminal nodes in the tree. Because the bracket keeps more structural
information of the original plan structure, which shows less ambiguity than
simple {\em BFS} and {\em DFS} strategies. The output of this strategy({\em
DFS-Bracket}) for our running example in Figure \ref{fig:plan_01} as shown in
\ref{table:tree_traversal}. We always sort the children nodes by typename, so
that the linearization of a tree is deterministic.

\begin{table}[tph]
    \begin{center}
    {\small
    \begin{tabular}{lp{6.6cm}} \toprule
    \bf{Strategy}                              & \bf{Node Sequence} \\ \midrule
    \multirow{3}{*}{\parbox{1cm}{DFS Bracket}} &
(Filter--, (Sort--, (Aggregate--, (Join-Hash-, (Loop--Nested, (Join-Hash-, (Hash--, (Loop--Nested, (Loop--Nested, Scan-Index-, Scan-Seq-) Scan-Heap-Bitmap) ) Scan-Index-Bitmap) Scan-Index-) Scan-Seq-))))                            \\ \bottomrule
    \end{tabular}
    }
    \end{center}
    \vspace{-5mm}
    \caption{\small Running examples for DFS-Bracket traversal
      Strategies. We use hyphens to connect 3 subtypes. When no
      sub-type for the node, we denote it as NIL type, here we use
      blank space for it to save table space. For example, the first
      node 'Filter-' actually means the first subtype is `Filter', the
      second and the third subtype is `NIL' }
\label{table:tree_traversal}
\vspace{-3mm}
\end{table}

\vspace{-2mm}
\Paragraph{Self-attentive Encoder Layer.} After we transform a tree-structure plan
into a sequence of normalized node types, various language modeling
architectures can be reused to learn encoding for the linearized plan.

Inspired by the success of attention mechanism in NLP, we employ the
multi-head, multi-hop attention mechanism used in Transformer
networks~\cite{vaswani2017attention} pictorially presented in Figure
\ref{fig:structure}. As before, due to space constraints, we refer the reader
to the original work for details. We will use the $(\bm{Q}, \bm{K}, \bm{V})$
notation from the original paper here. These matrices represent a query, key,
and value, respectively. The multi-head attention is defined as:
\begin{equation}
\label{eq:multihead_attention}
{\small \text{Multihead}(\bm{Q},\bm{K},\bm{V}) = [\text{head}_{1} \circ \ldots \circ \text{head}_{h}]\bm{W}^{O}}
\end{equation}
\begin{equation*}
 \text{head}_{i} = \text{softmax}\left(\frac{\bm{Q}\bm{W}^{Q}_{i}\left(\bm{K}\bm{W}^{K}_{i}\right)^T}{\sqrt{d}}\right)\bm{V}\bm{W}^{V}_{i}
\end{equation*}

The $\bm{W}_i$'s refer to projection matrices for the three inputs, and the
final $\bm{W}^o$ projects the concatenated heads into a single vector, and
$\frac{1}{\sqrt{d}}$ is scaling factor where $d$ is the dimension of $\bm{Q},
\bm{K},$ and $\bm{V}$. $\circ$ means concatenating the encoding attended by
multiple heads.

The choices of the query, key and value defines the attention mechanism.  In
our work, we use {\em self-attention}, defined by setting all three matrices to
$[\bm{n}_{j1}\ldots \bm{n}_{jk}]$, $\bm{n}_{jk}$ is the input encoding of the
$j$th self-attentive layer, which is corresponding the encoding of the $k$th
node in the linearized version, as described above.

\Paragraph{Input Embedding Layer}
In the $1$st layer, $\bm{n}_{11}$ is the input embedding of those operator
nodes. As mentioned above, for every operator node in the plan, we represent
its input embedding as a concatenation of embedding for its three subtypes.
Notice that in the {\em DFS-bracket} tree traversal, besides the regular nodes,
open and close bracket are also treated as nodes with three subtypes, which are
``BR\_OPEN-NIL-NIL" and ``BR\_CLOSE-NIL-NIL" for open and close bracket nodes,
respectively. When we use the self-attentive encoder, as the transformer usage
in BERT, we also add {\em CLS} node at the beginning and {\em SEP} node at the
end of the linearized nodes. They are also denoted as 3 subtypes as
``CLS-NIL-NIL" and ``SEP-NIL-NIL". Hence, in Table 2 Level 1 node, we actually
add four extra special subtypes: ``BR\_OPEN", ``BR\_CLOSE", ``CLS", ``SEP".

\Paragraph{Matching Layer}
The output of the transformer encoder is a sequence of vector for each nodes,
we use the output encoding of 'CLS' node as the encoding of the plan $p_{i}$,
because it aggregates the weighted sum of all other nodes in the self-attentive
layer.  We denote the plan encoding for $p_{i}$ as $P_{i} \in \mathbb{R}^{d}$.
After encoding the plan-pair $<p_{i}, p_{j}>$ into vectors $<P_{i} P_{j}>$,
then we use a matching layer to compute the similarity as
$$\sigma(W*[v_{i} \circ v_{j} \circ (v_{i}-v_{j}) \circ (v_{i}v_{j})]+b)$$

where $\sigma$ denotes the sigmoid activation function, $W \in
\mathbb{R}^{4d}$, $\circ$ means concatenating the four vectors.\footnote{Other
match function exists, e.g. bilinear similarity $v_{i}Mv_{j}^{T}$,
$M\in\mathbb{R}^{d \times d }$. However, we found that this contanated matching
similarity can largely reduce the parameters size from $d^{2}$ to $4d$ and
achieve better performance}


\subsection{Computational Performance Encoder}
\label{sec:computational}
In this section, we present our computational performance encoder, describing
the pretraining task to supervise the encoder learning, and our proposed model
architectures and the thoughts behind it.

\vspace{-1mm}
\subsubsection{Performance Attribute Prediction}
\label{sssec:computational_pretraining}
The properties mentioned in Table \ref{table:plan_prop} for each type of
broadly classified operators in a plan gives an ample hint on its computational
demand. These properties are either derived from complex logical inferences by
plan optimizers or real output from the query execution.  In previous works
\cite{ding2019ai, marcus2019plan, ding2018plan}, we notice the use of {\sf
Total Cost, Total Time, Startup Time} properties as a measure of performance.
We strongly agree with previous research works on using the properties
above-mentioned as measures of computational performance. Moreover, in our
encoder, we use these attributes as labels for prediction with an attempt to
encode the underlying features.  We use properties explicitly mentioned in
nodes (an instance of an operator in a plan), meta-information from databases,
and configuration settings of the database to predict these labels. In the
process of learning the labels, we learn the implicit features as embedding
with our computational performance encoder.

\begin{table}[t]
    \centering
    {\small
      \begin{tabular}{lp{\linewidth-2.8cm}} \toprule
        \bf{Features Type} &  \bf{Feature Attributes}                              \\ \midrule
        \multirow{2}{*}{Meta Features}   &  relname, attname, reltuples, relpages, relfilenode, relam, n\_distinct, distinct\_values, selectivity, avg\_width, correlation  \\ \hline
        \multirow{5}{*}{DB Settings}   &  bgwriter\_delay,  shared\_buffers, bgwriter\_lru\_maxpages, wal\_buffers, random\_page\_cost, bgwriter\_lru\_multiplier, checkpoint\_completion\_target, checkpoint\_timeout, cpu\_tuple\_cost, max\_stack\_depth, deadlock\_timeout, default\_statistics\_target, work\_mem  effective\_cache\_size, effective\_io\_concurrency, join\_collapse\_limit, from\_collapse\_limit,  maintenance\_work\_mem    \\ \bottomrule
      \end{tabular}
    }
  \vspace{-3mm}
  \caption{ Meta Features and DB Settings used as input features to Computational Performance Encoder}
   \label{table:featurenames}
    \vspace{-5mm}
\end{table}

We first create encoders, each for { (i) Scan (ii) Join (iii) Sort (iv)
Aggregate} functional operators, these four operators are the most frequently
used in query plans. The nodes with operator type {\sf Hash Join, Merge Join,
Nested Loop, Left/Right/ Inner/Outer Merge Join, Nested Loop} is mapped to
Join; similarly {\sf Seq Scan, Index Scan, Heap Scan, Bitmap Heap Scan} etc.
are Scan nodes.  From the properties of each node, we also extract the relation
names and attribute names of which it is accessing the data from node
properties such as {\sf Relation Name, Hash/Join/Merge/Index Condition, Filter,
Output}. We map them with the refreshed meta-information collected from the
database. In Table \ref{table:featurenames}, we show the meta-information
attributes we use as input to the model used by the node. This information can
be easily extracted from system tables of database system like PostgreSQL
\cite{stonebraker1990implementation}. In case of multiple relations used we sum
up the attribute values and then use them as input features.

We also use a set of database configuration setting values of the running
database as input features to the model. These configuration settings are
selected based on their importance in for performance tuning as described in
\cite{van2017automatic, site:postgistuning}.

Altogether, we have three types of input features,
\vspace{-1mm}
\begin{pkl}
\item {\em (a) Plan Features, $f_{node}$:} Explicitly obtained from a node of a plan, see Table \ref{table:plan_prop}.
\item {\em (b)  Meta Features, $f_{meta}$:} Meta-information about data and its distribution, see Table \ref{table:featurenames}.
\item {\em (c) DB Settings, $f_{db}$:} Handful number of database configuration settings, see Table \ref{table:featurenames}.
\end{pkl}
\vspace{-1mm}

To further clarify, the input features into a model are node properties
$f_{node}$ along with associated properties $f_{meta},f_{db}$ mentioned
earlier. For example, if a plan has three nodes $n_1,n_2,n_3$ of same operator
type, say Scan, then we have three input data $(f_{n_1},f_{meta},f_{db}) $,
$(f_{n_2},f_{meta},f_{db})$, $(f_{n_3}, f_{meta}$, $f_{db})$ for the scan
operator model. Besides, we create another input data with the summation of the
three node features of the plan $(f_{n_1}+f_{n_2}+f_{n_3},f_{meta},f_{db})$ for
the predicting the cumulative label metrics, i.e., {\sf Total Time} or {\sf
Total Cost} for the plan.

\subsubsection{Model Architecture.}
\label{sssec:performance_model}
\begin{figure}[t]
    \vspace{-1mm}
    \centering
    \includegraphics[width=0.48\textwidth]{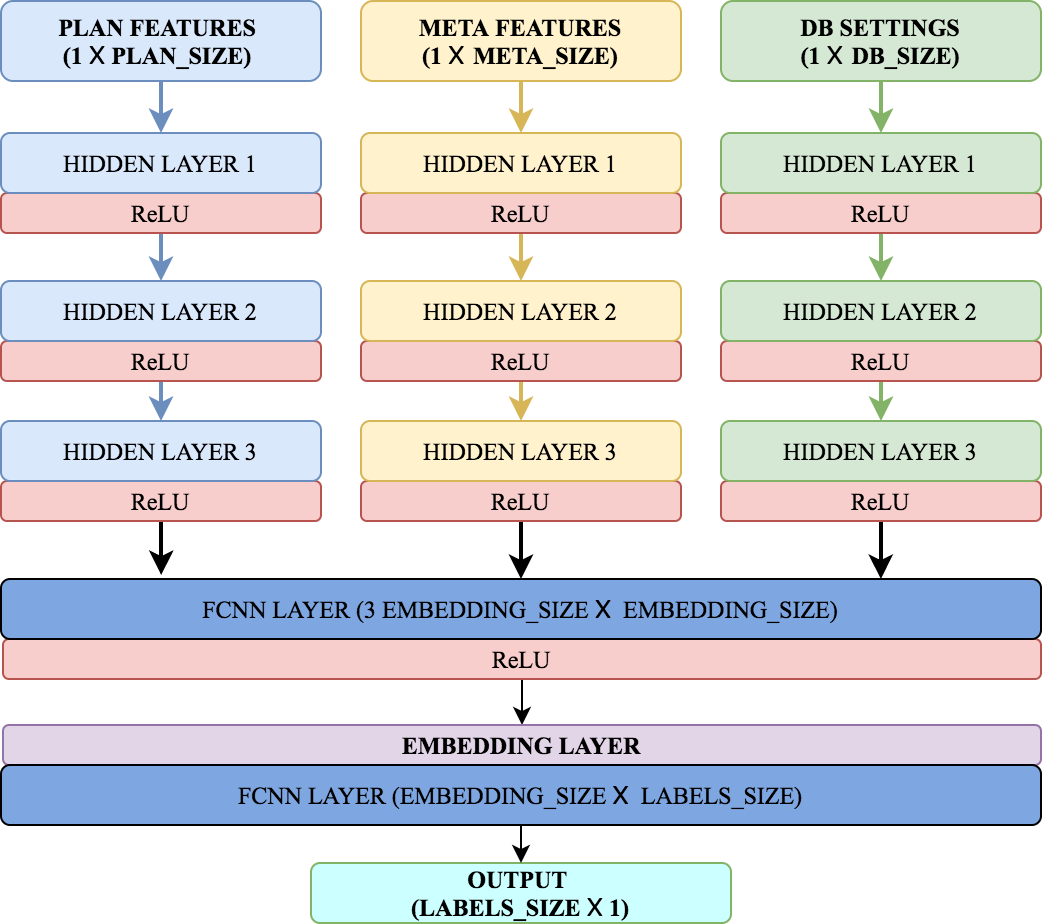}
    \captionsetup{justification=centering}
    \vspace{-6mm}
    \captionof{figure}{The multicolumn deep neural network(DNN) for our computational performance encoder.}
    \label{fig:performance}
    \vspace{-6mm}
\end{figure}

We now present the deep neural network (DNN) architecture of the encoder with a
pictorial representation in Figure \ref{fig:performance}. It is a three-column
DNN on the top each for Plan features, Meta features, and DB features,
respectively, with another fully-connected NN layer merging the three parts and
producing the embedding layer. The last fully-connected NN component takes the
output of the embedding layer to predict the metric labels i.e., {\sf Total
Cost, Total Time, Startup Time}. Also, each NN layer is followed by a
activation function layer of ReLU (Rectified Linear Unit), Sigmoid or Tanh
functions. We create multiple instance of this supervised regression model each
for a type of functional operator as mentioned earlier.

\begin{figure*}[!t]
    \vspace{-3mm}
    \centering
    \includegraphics[width=\textwidth]{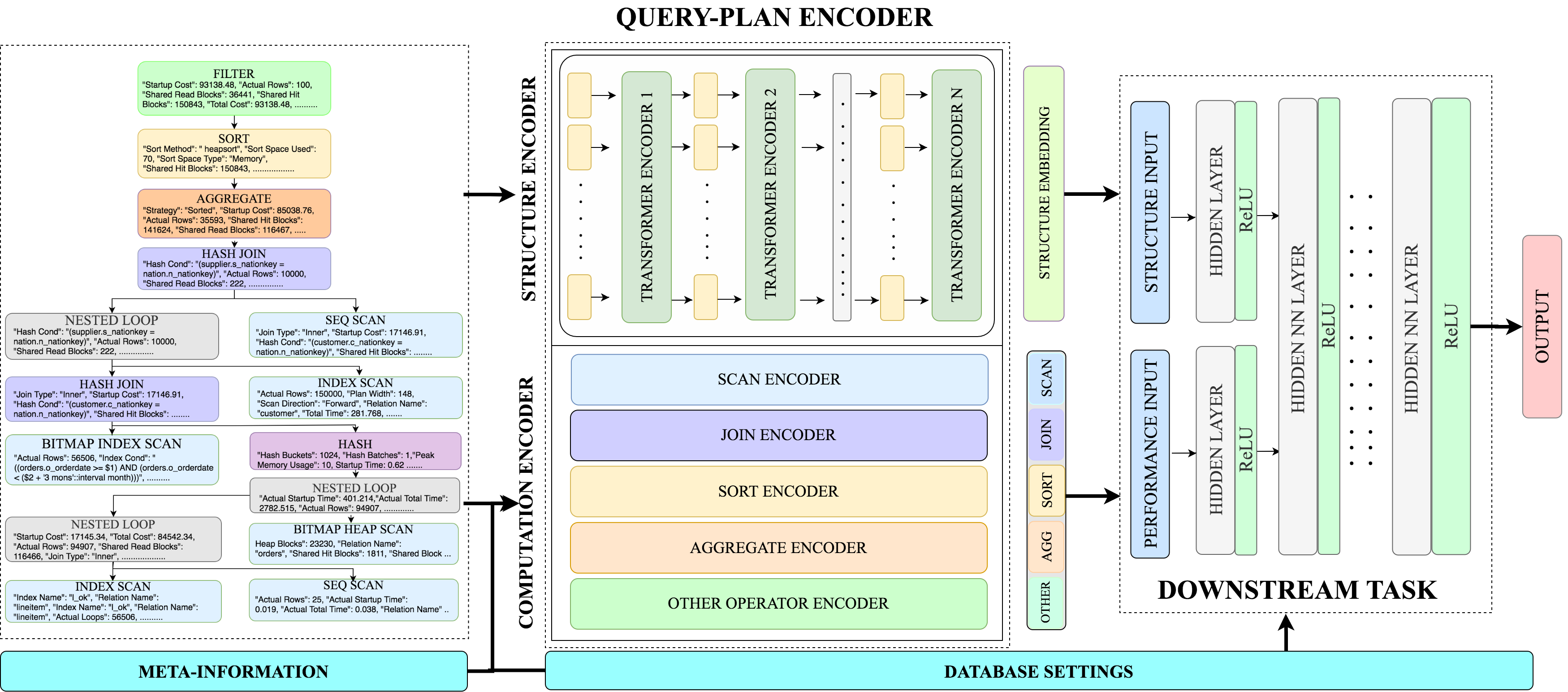}
    \captionsetup{justification=centering}
    \vspace{-6mm}
    \captionof{figure}{A bird-view diagram, showing the role of plan encoders for downstream task.}
    \label{fig:overall}
    \vspace{-5mm}
\end{figure*}

A NN layer can efficiently represent or capture complex relations among input
features by applying an affine transformation of the input. With multiple
feed-forward NN layer, a recursive affine transformation and non-linear
activation functions are applied to the input features to produce an output.
The difference between the desired output and the predicted output is
calculated based on some metric functions dubbed as loss. A gradient descent
based technique is applied to tweak the weights on each layer used to perform
the affine transformation minimizing the loss. It allows the model to learn
non-linear and polynomial order complex functions, automatically identifying
the relevant features.

One of the key insights while architecting the model is that the three-column
multilayered feature approach on Plan, Meta, and DB features, respectively,
allows the model to find correlation among the same type of features first.
Then transformed weighted features from each part can correlate effectively.
As a preliminary attempt, we train an alternate model with a standard
(single-column) DNN with all the input features together. In
\S\ref{sec:exp_computation}, we provide a comparative study to evaluate both
the models.

\subsubsection{Joint Training}
\label{sssec:joint_training}

A general rule of thumb for any model is that the distribution of predicted
data remains the same as training data. But, in our case, the data
distributions change with new workload. When the model learns from a single or
small workload benchmark, the model overfits and cannot generate a general
model. With the assumption that if enough information on the data distribution
is used for training the model, the model can learn the factors governing the
performance metrics for each operator (Scan, Join, Sort, Aggregate, etc.).
Also, the fact that a general query plan optimizer (which is a logical
component) uses the same statistical information we use as input to our model
encourages us. The trick is to learn a generalized pretrained model that can
adapt to an unseen workload with small data from the new domain. Hence, the
pretrained models should utilize already learned parameters to adapt with the
new workload.

We utilize a {\em joint training} approach for training the encoders.  We train
each operator model on multiple workloads on different data distributions and
multiple database configuration settings. In joint training approach, we
perform multiple metric tasks, each task optimizes for each label, i.e. {\sf
Total Cost}, {\sf Total Time} and {\sf Startup Time}. The difference in each of
these models is the last NN-layer, which uses the embedding layer as input.
Since the top level of the model remains unchanged, the weights are naturally
tweaked to learn features based on multiple tasks.

We evaluate our performance encoder models on two criteria, (i) the model uses
less data from a new domain to adapt, and (ii) the model error on validation
and test data converges. We provide a detailed evaluation results on our
pretrained computational performance encoder in \S\ref{sec:experiments}.

\vspace{-2mm}
\subsection{Finetuning Evaluation}
\label{ssec:finetuning}

Given the above pretraining for learning structure and computational
performance encoders, we hope that our learned model can be easily used in
other unseen applications. We conduct two groups of finetuning evaluation for
them:

\Paragraph{Domain Adaptation.}  For both the structure encoder and
computational encoder, they are trained from a source distribution on plan-pair
similarity regression and performance attribution prediction tasks.  {\em
Domain Adaptation} aims at that these models can be easily finetuned on a
different target data distribution. Hence, we finetuning them on different
benchmark workloads on the same tasks, such as TPC-H and TPC-DS , and Spatial
benchmarks. For plan-pair similarity regression task, we generate a collection
of plan pairs for each new benchmark, and then calculating the {\em Smatch}
scores for evaluation. For the performance attribute prediction task, we
collect the new dataset by running workloads on different database
configurations. More details about those datasets is introduced in
\ref{sec:datasets}, and the results on domain adaptation for each encoder are
shown in \S\ref{sec:analysis}.

\Paragraph{Transfer Learning to New Tasks} Besides the ability of domain
adaptation, we also define two new tasks to evaluate whether our pretrained
plan encoder can be easily used for other tasks rather than our pretraining
task in \S\ref{sec:downstream_tasks}.


\vspace{-1mm}
\section{Downstream Tasks}
\label{sec:downstream_tasks} In this section, we show two downstream tasks that
use our proposed plan structure and performance encoders. We present a bird-eye
view architecture of the model common to our downstream tasks in Figure
\ref{fig:overall}. For a given query plan input, meta information of database,
and database configuration, the plan encoders (structure and performance
encoder) produce respective representations as output. This output is then fed
to the downstream task-specific model. It is to note that for generating the
computational performance representation, we group plan nodes based on the type
of functional operator and then pass it to the corresponding performance
encoder to obtain representation.

The downstream task model is a standard multilayer-DNN taking three inputs, (a)
structure embedding,(b) computational performance embedding, and (c) the
database settings. The properties of database settings are real numbers. They
can have an arbitrarily large value, which hinders learning a better model. We
overcome the problem by scaling each database settings with logarithmic
function and use them as added features along with the real numbers.
Furthermore, we added a flexible design of reshaping the dimension of structure
or performance representation in the downstream task model for obtaining better
accuracy.

\subsection{Query Latency Prediction}
\label{sec:latency_prediction}
The first downstream task is a real-world task of predicting query latency for
an input query plan on a given database knob configuration settings utilizing
our plan encoders. Formally, we define the query latency prediction problem as
follows.

\vspace{-2mm}
\begin{problem}[Query Latency Prediction:] Given a query plan $p$,
meta-features $f_{meta}$ of the database, and a database configuration settings
$f_{db}$, the model predicts the latency of the query.  
\end{problem}
\vspace{-2mm}

For generating the training data for latency model, we created an automated
workload running scripts\footnote{\url{https://github.com/debjyoti385/workload_scripts}} that
runs on cloud server instances and uploads executed plans along with the
meta-features and database settings to our data repository.  The script
generates a new database configuration and configures the database
automatically for each run. These new database configuration are generated
based on the Latin Hypercube Sampling method \cite{audze1977new,
mckay2000comparison} for the properties mentioned in Table
\ref{table:featurenames}. This method for generating database setting has been
earlier used by Duan et al. and Aken et al. \cite{van2017automatic,
duan2009tuning}.

\subsection{Query Classification}
\label{sec:query_template}
One of the aims of workload characterization is to learn the features of
similar queries to classify and cluster them, thus providing an opportunity for
obtaining database instance optimality. This task also validates the efficient
representation of similar queries from our encoders by projecting them closely
in latent dimensions. Besides the latency prediction task, we also conduct
experiments on query template prediction task with our pretrained plan
structure and performance encoders. We formally define the problem statement as
follows.

\vspace{-2mm}
\begin{problem}[Query Classification:] Given a query plan $p$, meta-features
$f_{meta}$ of the database, and a database configuration settings $f_{db}$, the
model predicts the predefined class for the query plan based on feature
similarities.  
\end{problem}
\vspace{-2mm}

We conduct this experimental task with join order benchmark \cite{leis2015good}
containing 113 interesting query templates and 33 clusters of similar query
templates. Due to cardinality of the database tables and query predicates, the
query plans generated from the query optimizers can differ from one another. It
also makes the classification task challenging to cluster the query features
accordingly. Note that we include the performance encoder in classification
tasks as queries even with similar plan structure can differ in performance
features. We present detail of this experiment and the role of individual
encoders in \S~\ref{ssec:query_template_exp}.


\vspace{-2mm}
\section{Experiments and Results}
\label{sec:experiments}
In this section we first describe the datasets we used in our
experiments. We then present evaluation methods with experimental
results for latency prediction and query template classification tasks.

\vspace{-1mm}
\subsection{Datasets}
\label{sec:datasets}
\Paragraph{Crowdsourced Plan Dataset.}  We collected this dataset containing
PostgreSQL queries along with its execution plans from a crowdsourced
website\footnote{\url{https://explain.depesz.com}}\cite{site:depesz}. We used
this dataset for pretraining our structure encoder model. After pruning the
plans with more than 200 nodes, we generate 57430/2871/2871 plan-pairs for
training/dev/test, and then caculate the Smatch score as their similarity
score.

\vspace{-1mm}
\Paragraph{Industry Standard Benchmarks.} We have used two industry-standard
TPC-H \cite{bench:tpch} and TPC-DS \cite{bench:tpcds} benchmarks as workloads
with different scale factors, and execute them with different database
settings.

\Paragraph{Spatial Benchmark.} Spatial queries are notorious for hogging
resources and needs a proper database configuration for optimal performance.
PostGIS, the spatial and geographic objects extension for PostgreSQL admits the
configuration tuning requirement based on workload type in their documentation
\cite{site:postgistuning}. We use the two following spatial benchmarks in our
experiments.

{\em Jackpine:} Jackpine \cite{ray2011jackpine} benchmark contains diverse
spatial queries on spatial join with multipolygons, lines, points and
combination of them. We
revised\footnote{\url{https://github.com/debjyoti385/jackpine}} the original
benchmark with recently available shape datafiles, PostGIS extension and also
made it publicly available.

{\em Open Street Map (OSM):} The Open Street Map(OSM) workload has spatial
overlap, distance and routing queries. This dataset is
created\footnote{\url{https://github.com/debjyoti385/osm_benchmark}} with
inspiration from work \cite{baas2012nosql}. Due to sparsity, it is difficult to
understand the underlying data distribution, which makes it an inviting
benchmark for experiment. We used OSM map of New York and Los Angeles county.

\vspace{-1mm}
\Paragraph{Join Order Benchmark.} It contains 113 different queries, which can
be grouped into 33 clusters due the the similar SQL queries with different join
orders. We run those queries on different database configurations and then
collect the 16229 diffrent plans. We split that into 13505, 1362, 1362 as
training, dev and test respectively.


\vspace{-3mm}
\subsection{Results on Query Latency Prediction}
\label{sec:latency_pred_exp}
We first evaluate our query latency prediction model with multiple experiments
to project an overall effectiveness of using our plan encoders. We used
pretrained structure and performance plan encoders trained on Crowdsourced
dataset and multiple TPC-H, TPC-DS workloads, respectively. A detailed analysis
of our pretrained encoders is given in \S\ref{ssec:struct_exp} and
\ref{sec:exp_computation}.

\vspace{-1mm}
\Paragraph{Ablation Studies.}
{\em  (a) {Spatial Benchmark:}}
We first present an ablation study on individual queries. The aim of this study
is to measure the error in relative to the variability of query latency.  For
initial training of the latency prediction model, we used plans from spatial
benchmark \cite{ray2011jackpine,OpenStreetMap, barabasi2005origin} executed on
120 different database configurations. The trained model then predicts query
latency for spatial queries on different database configuration. To prepare our
test datasets, we ran each benchmark 50 times with very different database
configuration settings.

\begin{table}[!t]
    \begin{adjustbox}{width=1.02\linewidth,center}
    \begin{tabular}{lcrrr} \toprule
    \textbf{Database Setting} & \textbf{Unit} & \textbf{Median} & \textbf{95th Percentile} & \textbf{5th percentile} \\ \midrule
    bgwriter\_delay           &	ms            &	4,860.00	& 9,421.05                 & 456.00                  \\
bgwriter\_lru\_maxpages       &	integer       &	515.00          & 958.05                   & 55.00                   \\
checkpoint\_timeout           &	ms            &	300.00          & 540.00                   & 60.00                   \\
deadlock\_timeout             &	ms            &	300,000.00	& 540,000.00               & 26,000.00               \\
default\_statistics\_target   &	integer       &	4,827.50	& 9,563.00                 & 454.85                  \\
effective\_cache\_size        &	bytes         &	1,048,576.00	& 1,966,080.00             & 131,072.00              \\
effective\_io\_concurrency    &	integer       &	52.00           & 96.00                    & 6.00                    \\
maintenance\_work\_mem        &	bytes         &	7,340,032.00	& 15,728,640.00            & 876,953.60              \\
max\_stack\_depth             &	integer       &	3,072.00	& 5,120.00                 & 417.95                  \\
random\_page\_cost            &	number        &	5,028.60	& 9,507.39                 & 560.40                  \\
shared\_buffers               &	bytes         &	2,097,152.00	& 3,932,160.00             & 131,072.00              \\
wal\_buffers                  &	bytes         &	130,624.00	& 131,072.00               & 12,416.00               \\
work\_mem                     &	bytes         &	15,728,640.00	& 31,457,280.00            & 1,048,576.00            \\ \bottomrule
    \end{tabular}
\end{adjustbox}
\vspace{-2mm}
\caption{\footnotesize Statistics on configuration settings generated for training data.}
\label{table:config_statistics}
\vspace{-5mm}
\end{table}

\begin{figure}[t]
    \centering
        \includegraphics[width=\linewidth]{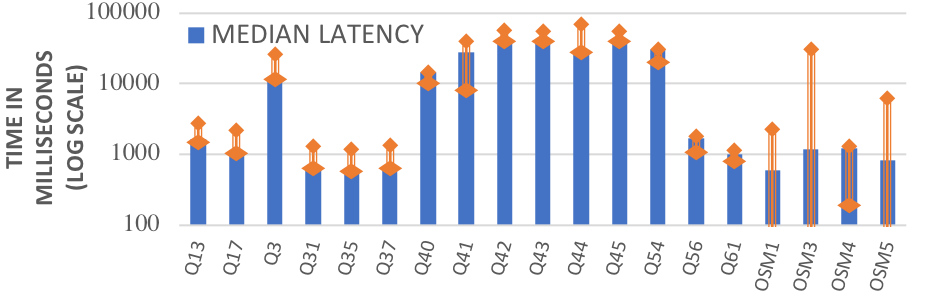}
        \vspace{-7mm}
        \caption{\footnotesize Statistics on latency of spatial queries ($>500$  milliseconds) from Jackpine \cite{ray2011jackpine} and OSM benchmark, where the blue bar represents median, the orange line represents the {\em variability} with 5th and 95th percentile of query latency for different database configuration.}
        \label{fig:latency_spatial}
        \vspace{-3mm}
\end{figure}

\begin{figure}[!t]
    \centering
        \includegraphics[width=\linewidth]{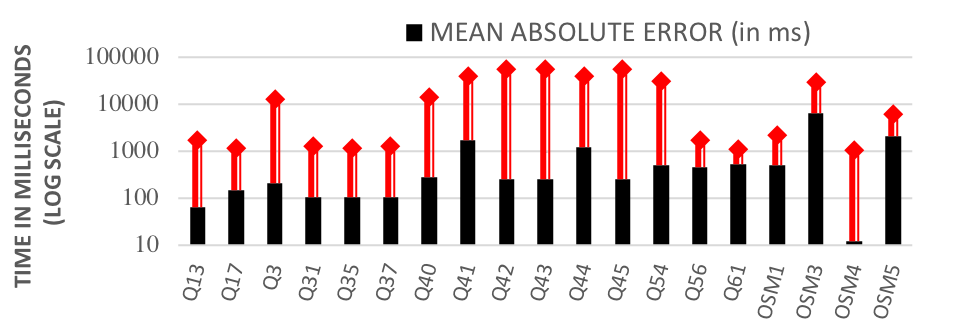}
        \vspace{-7mm}
        \caption{\footnotesize The black bar represents mean absolute error (MAE) (in milliseconds) for spatial Jackpine and OSM queries, the red line represents the {\em variability} i.e. measure of time difference between 95th percentile and 5th percentile (same as the orange line from Figure \ref{fig:latency_spatial}).}
        \label{fig:mae_spatial}
        \vspace{-3mm}
\end{figure}

Figure \ref{fig:latency_spatial} shows the query latency statistics of query
templates with median query latency greater than 500 milliseconds from spatial
benchmark; Jackpine (with prefix Q) and OSM benchmark (with prefix OSM). The
blue bars in the chart shows the median of the query latency for all the query
execution with different database settings. The orange line shows the
variability of the query latency due to change of database settings. The bottom
point of the orange line represents the 5th percentile, and the highest point
marks the 95th percentile of query latency.  We present a complimentary Figure
\ref{fig:mae_spatial} along with Figure \ref{fig:latency_spatial} that
pictorially shows the mean absolute error for all the query templates from the
spatial benchmark. The red line is the measure of time difference between 95th
percentile and 5th percentile of a query latency in milliseconds, depicting the
extent of the {\em variability} for the particular query. To note, vertical
axes on both figures i.e. Figure  \ref{fig:latency_spatial} and
\ref{fig:mae_spatial} are presented on a logarithmic scale with milliseconds as
unit. It shows that at least 68\% of the queries have MAE less than 10\% of
{\em variability}, and 90\% of the queries have MAE less than 30\% of {\em
variability}.


\begin{figure*}[t]
    \centering
        \includegraphics[width=\textwidth]{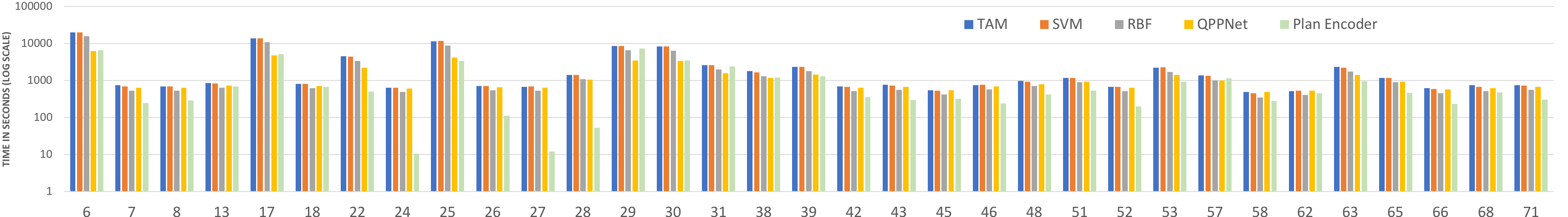}
        \vspace{-5mm}
        \caption{\footnotesize Mean absolute error (MAE) for the 33 TPC-DS query templates with scale factor 100 where Plan Encoder performed better than a baseline.}
        \label{fig:latency_tpcds100_good}
        \vspace{-3mm}
\end{figure*}
\begin{figure*}[!t]
    \centering
        \includegraphics[width=\textwidth]{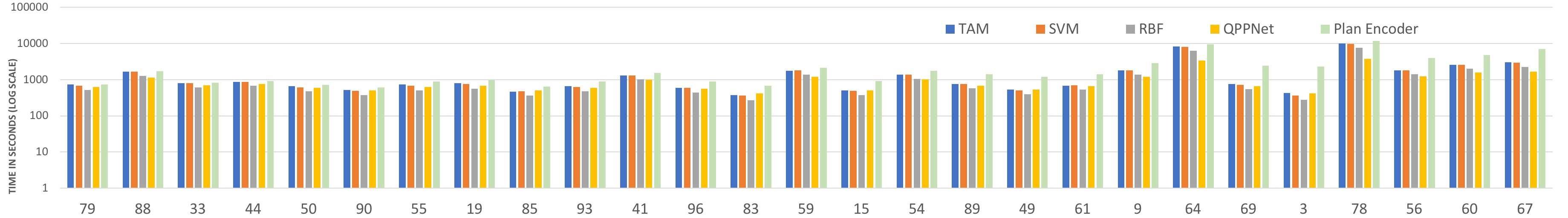}
        \vspace{-7mm}
        \caption{\footnotesize Mean absolute error (MAE) for the 27 TPC-DS query  templates with scale factor 100 where Plan Encoder did not performed better than a baseline.}
        \label{fig:latency_tpcds100_bad}
        \vspace{-5mm}
\end{figure*}

Query latency prediction on spatial benchmark is challenging because of the
sparse geospatial data distribution from two areas contributing towards large
variability. Furthermore, the performance of spatial queries are easily
affected by database configurations. Significantly less mean absolute error
from the latency prediction model shows that pretrained encoders helped the
model.

{\em  (b) {TPC-DS 100 Benchmark:}} In this experiment, we compare our latency
model with state-of-the-art latency prediction models for each query template
from TPC-DS benchmark for a scale factor of 100 (i.e. 100 GB). A recent study
by Marcus et al.\cite{marcus2019plan} shows TPC-DS query ablation study with
{\sf TAM} \cite{wu2013predicting}, {\sf SVM} \cite{akdere2012learning}, {\sf
RBF} \cite{li2012robust} and {\sf QPP Net} \cite{marcus2019plan}. It is to note
that we used the same TPC-DS plan dataset used by the study
\cite{marcus2019plan}, and we split of our dataset in  80:20 ratio for use as
training and test data.  In Figure \ref{fig:latency_tpcds100_good} and
\ref{fig:latency_tpcds100_bad}, we show a comparative study of our latency
model with other approaches on mean absolute error (MAE) for each TPC-DS query
template.

Figure \ref{fig:latency_tpcds100_good} shows all the query templates where our
latency model with plan encoders performed better than the majority of the
baseline approaches. 23 out of 33 query templates produced at least 25\% less
error than the best baseline for respective query templates.

In Figure \ref{fig:latency_tpcds100_bad}, we present all the query templates
where our latency model could not perform better than the baselines. 12 out of
27 query templates achieved a mean error of less than 25\% of the baseline. 21
out of 27 templates have errors less than twice of the baseline. We noticed
that for some query templates, there is a large gap in latency prediction for
non-indexed versions of the queries from the indexed version of the queries. We
could not collect enough metadata information for this particular dataset on
indexed columns to fully replicate the database condition and then to use it in
our performance plan encoder. We used metadata information for non-indexed
database configuration for TPC-DS with scale factor 100 in our plan encoder for
this prediction. Thus from the experiments, it shows that our initial approach
of plan encoders can be used for latency prediction strategies.

\begin{figure}[t]
    \centering
        \includegraphics[width=\linewidth]{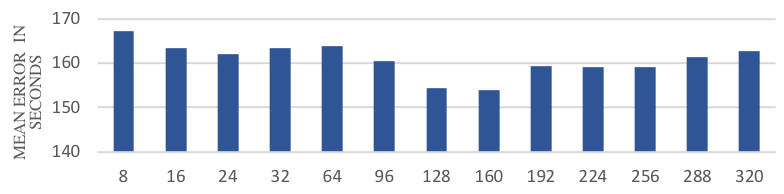}
        \vspace{-7mm}
        \caption{\footnotesize Average of MAEs on 5 test datasets of TPC-DS with scale factor 10 with varying embedding size of structure encoder.}
        \label{fig:str_vs_perf}
        \vspace{-5mm}
\end{figure}

{\em Discussion on Structure Encoder Size:}
We performed an experiment to find the optimal size of the structure encoder to
be used in the latency model. In latency prediction, we find using only
structure encoder yields error 5 times of the latency model with only
performance encoder. We designed an experiment to show the influence of
structure encoder on the latency prediction model by varying the input
embedding size from structure encoder and keeping the performance embedding
size fixed to 300. We trained the latency model with TPC-DS scale factor 10
dataset and tested it on 5 different test batch of TPC-DS dataset. In Figure
\ref{fig:str_vs_perf}, we report the average of the mean absolute error from 5
test dataset and found that embedding size of 128 and 160 performed relatively
well. It signifies that adding structure can help latency prediction by a small
amount with a suitable structure embedding size.  This confirms that features
from performance encoder are dominant and have relatively low importance of
structure features in latency prediction task.

\subsection{Results on Query Classification}
\label{ssec:query_template_exp}
As described in \S\ref{sec:query_template}, our experiments are conducted on
join-order benchmark, we fuse our pretrained structure encoder and performance
encoder to classify a plan into a template id.  Noticed that in the join order
benchmark, there are 113 query templates and 33 clusters. We hope our
classifier can consistently predict both the cluster id and template id. For
example, when predicting a plan with ground truth template id {\sf 11a}. It
means the ground truth cluster id is {\sf 11}. In this cluster, we also have
another three template id {\sf 11b, 11c} and {\sf 11d}. We sum up all 4 scores
of those 4 template id in a cluster as the score for predicting the cluster id
as 11. Then we add another cross entropy score for the clustering prediction as
a regularizer of our template id cross entropy. Adding this regularizer largely
improves the performance of our model. Besides that, in this task, we found
adding a batch normalization layer when fusing our structure and performance
encoder as inputs is essential, which normalizes them into the same scale. To
understand how structure and performance encoder performs in the task, we
conduct ablation studies for using structure-only, performance-only and both in
our experiments. The results in Table \ref{table:rst_query_classication} show
structure encoder plays in the main role of this task. Without it, the
performance-only performs very bad. What's more, adding the performance encoder
can boost the performance by 5.72 points on the test set. We also noticed that
using both structure and performance only generalize better on unseen test set
than structure-only or performance-only models. Below the line in the table,
{\em Both0.1} and {\em Both0.3} are the models only trained on 0.1 and 0.3
fraction of data, they perform still good even in less amount of data, which
indicates that our pretrained encoder help on other new tasks with less data.

\begin{table}[h]
\vspace{-2mm}
\begin{center}{\small
\setlength{\tabcolsep}{3pt}
\begin{tabular}{lllll} \toprule
 \multirow{2}{*}{Methods} & \multicolumn{2}{c}{Dev} & \multicolumn{2}{c}{Test}                   \\ \cmidrule(lr){2-3} \cmidrule(lr){4-5}
                          & template                & cluster      & template     & cluster      \\\midrule
Structure-Only            & 0.2452                  & 0.4670       & 0.1946       & 0.3847       \\
Performance-Only          & 0.1645                  & 0.2973       & 0.0977       & 0.1769       \\
Both                      & {\bf 0.2783}            & {\bf 0.5573} & {\bf 0.2518} & {\bf 0.4647} \\\midrule
Both0.1                   & 0.2000                  & 0.4927       & 0.151        & 0.334        \\
Both0.3                   & 0.2555                  & 0.5228       & 0.1843       & 0.3855       \\\bottomrule
\end{tabular}}
\end{center}
\vspace{-5mm}
\caption{\label{table:rst_query_classification} Results on Query Classication Accuracy}
\vspace{-4mm}
\end{table}


\vspace{-2mm}
\section{Analysis}
\label{sec:analysis}

\vspace{-1mm}
\subsection{Structure Encoder}
\label{ssec:struct_exp}
As described in \S\ref{sec:structure}, our structure encoder is pretrained on
plan-pair similarity regression task with the transformer encoder. We use a
large amount of dataset from the Crowdsourced Plan dataset for pretraining. In
this paper, we first prune those extremely large plans with more than 200
nodes. Then randomly select 63172 pairs of plans to form the dataset for our
plan-pair regression task and calculating all the smatch scores of those pairs.

\Paragraph{Baseline Models}
For pretraining tasks, we denote our plan-pair similarity regression task as
{\em Transformer-PPSR}, and compare it a common self-supervised pretraining
tasks: Sparse Autoencoder({\em Sparse-AE}). It learns to compress a input plan
into a hidden representation and then decode that representation back into the
original plan. Besides that we also conduct a blation study for using LSTM on
our PPSR pretraining task, denoted as {\em LSTM-PPSR}.

\begin{figure}[t]
    \vspace{-3mm}
    \includegraphics[width=\linewidth]{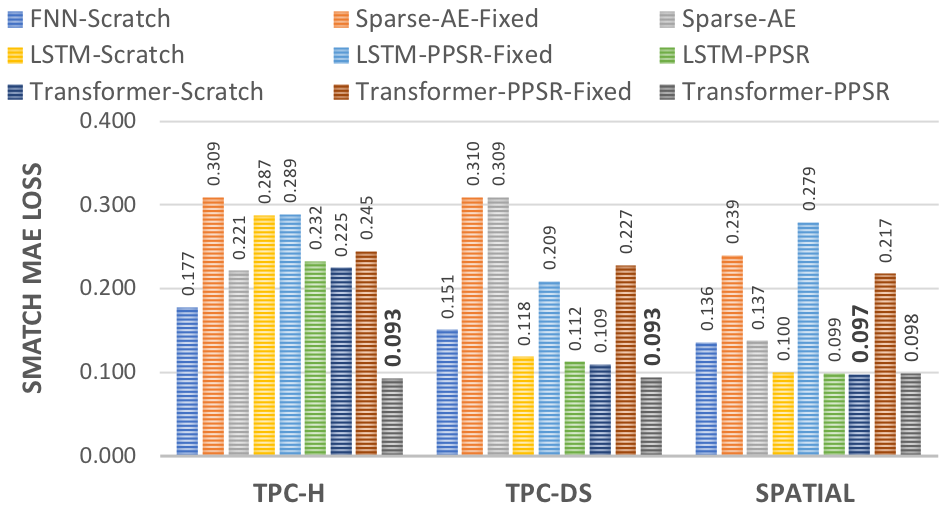}
    \vspace{-7mm}
    \caption{Main Results of finetuning structure encoder on TPC-H, TPC-DS, and SPATIAL}
    \label{fig:structure_results}
    \vspace{-7mm}
\end{figure}

\Paragraph{Results on Finetuning}
We first pretraining on Crowdsource dataset with 3 pretraining approaches: {\em
Sparse AE}, {\em LSTM-PPSR}, {\em Transformer-PPSR}. To investigate its ability
for domain adaptation, we first random generate 11126, 55498, 60000 plan-pairs
with plans in TPC-H, TPC-DS, and SPATIAL, then creating the training, dev, test
splits with a ratio as $20:1:1$.  We finetuning the pretrained structure
encoder on the above new domains. In Figure \ref{fig:structure_results}, we
show the Mean Absolute Error (MAE) between the predicted smatch score with the
true smatch score.  All the models in the table are trained from the full
dataset. As described above, the methods above the line are training from
scratch with FNN, LSTM, Transformer Encoder.  The models below the lines are
the 3 pretraining methods, and we use each of them as fixed feature or
finetuning.  `Transformer-PPSR-fixed' means using the output of the pretrained
model as fixed features without finetuning, which performs much worse than its
finetuning version `Transformer-PPSR'. Hence, our models are suitable for
finetuning instead of fixed features in this task. Except for the MAE on the
spatial dataset, our pretrained model, can significantly reduce the error on
both TPC-H and TPC-DS. For spatial dataset, we noticed that both LSTM and
Transformer scratch models work very good; using pretraining does not improve.

\begin{figure}[t]
    \vspace{-3mm}
    \includegraphics[width=\linewidth, height=4.5cm]{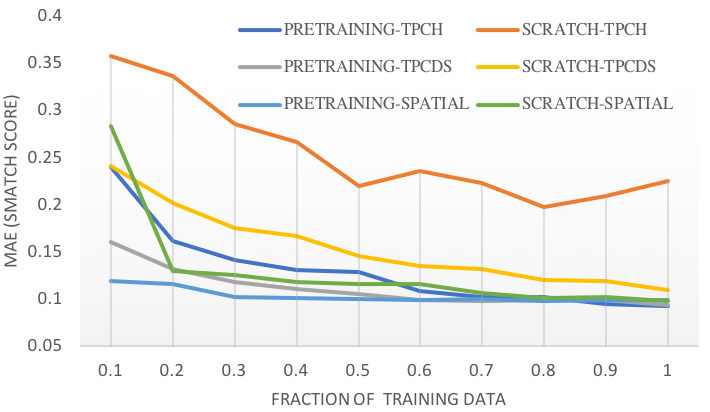}
    \vspace{-7mm}
    \caption{Plan-pair Regression: MAEs of Smatch score on fractions of training data}
    \label{fig:smatch_mae}
    \vspace{-3mm}
\end{figure}

As shown in the Figure \ref{fig:smatch_mae}, we compare pretraining
and no-pretraining method with different amount of training data. For
all 3 benchmarks, especially TPCH and TPCDS, our pretrained method can
achieve small MAE of Smatch score on less amount of data. On spatial
data, our pretrained method only slightly better than no-pretraining
one.

\vspace{-2mm}
\subsection{Computational Performance Encoder}
\label{sec:exp_computation}
We now perform local probe on computational performance encoder with a set of
experiments evaluating the pretrained encoders for Scan, Join, Sort, and
Aggregate operator. For pretraining, we used TPC-H and TPC-DS both with scale
factors 1,2,3 and 5 executed on at least 20 different configuration settings
randomly generated via Latin Hypercube Sampling method \cite{audze1977new,
mckay2000comparison}.

\begin{figure}[t]
    \centering
    \begin{subfigure}[b]{0.48\linewidth}
        \includegraphics[width=\linewidth]{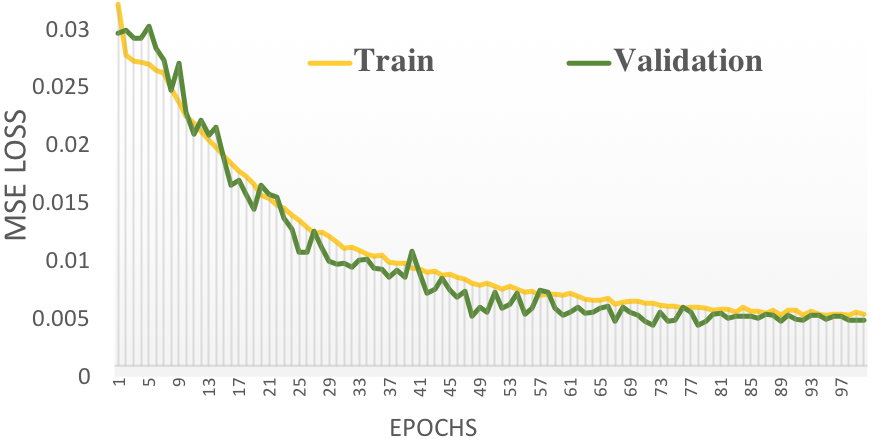}
        \vspace{-5mm}
        \caption{\small Structure encoder training.}
        \label{fig:structure_pretraining}
    \end{subfigure}
    \begin{subfigure}[b]{0.48\linewidth}
        \includegraphics[width=\linewidth]{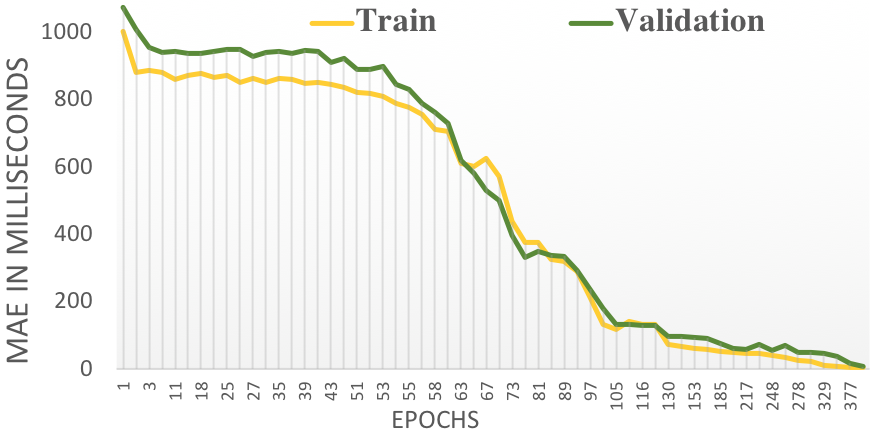}
        \vspace{-5mm}
        \caption{\small Scan operator pretraining.}
        \label{fig:scan_pretraining}
    \end{subfigure}
    \begin{subfigure}[b]{0.48\linewidth}
        \includegraphics[width=\linewidth]{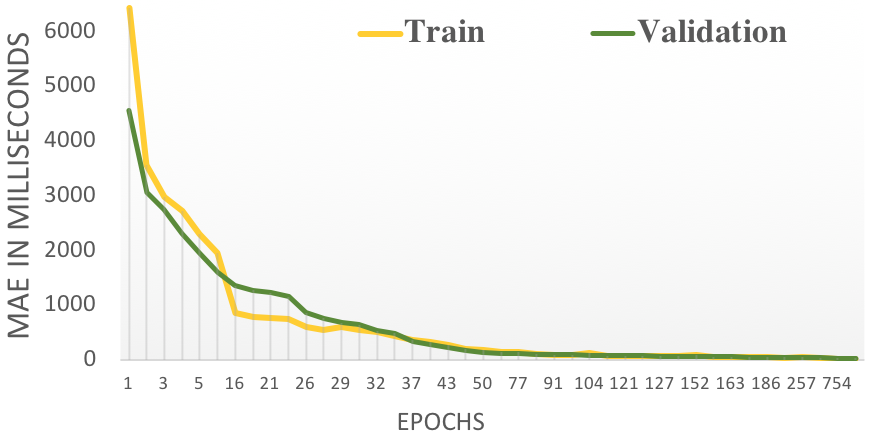}
        \vspace{-5mm}
        \caption{\small Join operator pretraining.}
        \label{fig:join_pretraining}
        \vspace{-2mm}
    \end{subfigure}
    \begin{subfigure}[b]{0.48\linewidth}
        \includegraphics[width=\linewidth]{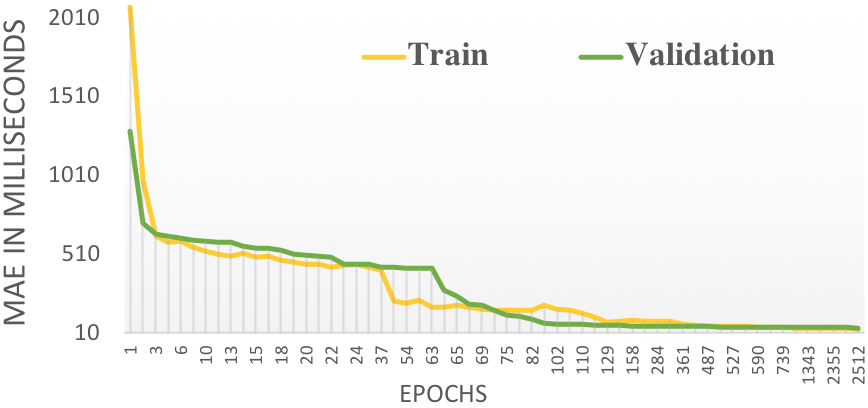}
        \vspace{-5mm}
        \caption{\small Sort operator pretraining.}
        \label{fig:sort_pretraining}
        \vspace{-2mm}
    \end{subfigure}
    \vspace{-1mm}
    \caption{\small Showing convergence of Mean absolute errors(MAE) (in seconds) for the validation, test and train datasets, while pretraining all the computational performance encoders.}\label{fig:operators_pretraining}
    \vspace{-3mm}
\end{figure}

\Paragraph{Pretraining:}
We first illustrate the training procedure and a few learnings from it. We
split the dataset into 8:1:1 ratio for train, validation, and test for
pretraining of all the four operators. Figure \ref{fig:operators_pretraining}
shows the Mean Absolute Error (MAE) on latency ({\sf Actual Total Time}) label
for train, validation, and test data for scan, join, and sort operator. In all
the cases along with aggregate (not shown in Figure
\ref{fig:operators_pretraining}) the train, validation and test MAE converges
below 1 second and stays around tens of milliseconds.  The MAE on test data is
calculated based on the epoch with best validation model so far seen while
training.  We stop the training when the MAE on validation does not improve
more than 5 milliseconds in last 100 epochs.  With a 12 GB GPU on a Ubuntu
18.04 operating system, each model takes around 6-8 hours to train.

A key insight on training the models is the best MAE vary based on operators.
The best MAE for Scan model on test data is 12 milliseconds, where the
validation MAE is 7 milliseconds. In Join model and Sort model the test MAEs
reaches a low of 3.42 milliseconds and 44 milliseconds respectively.  It is to
note that we performed pretraining on all the three labels {\sf Actual Total
Time, Total Cost} and {\sf Startup Time} but for brevity we reported only {\sf
Actual Total Time} in our figures.

\begin{figure}[t]
    \centering
    \begin{subfigure}[b]{0.48\linewidth}
        \includegraphics[width=\linewidth]{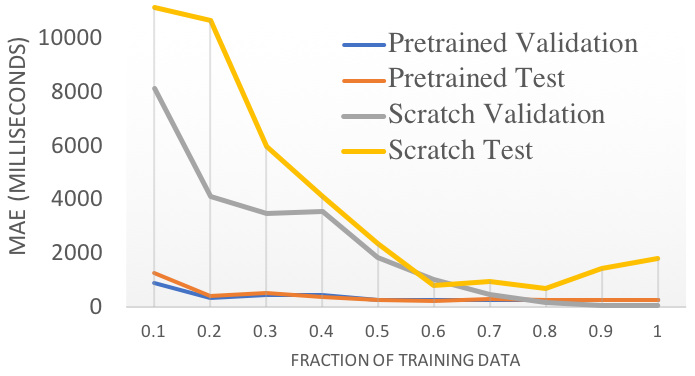}
        \vspace{-5mm}
        \caption{\small MAEs on Scan operator model for fractions of training data on TPC-DS SF-8.}
        \label{fig:scan_finetuning_tpcds}
    \end{subfigure}
    \begin{subfigure}[b]{0.48\linewidth}
        \includegraphics[width=\linewidth]{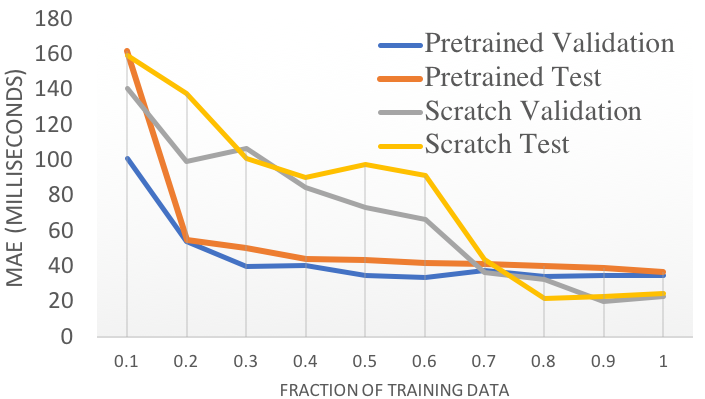}
        \vspace{-5mm}
        \caption{\small MAEs on Join operator model for fractions of training data on TPC-DS SF-8.}
        \label{fig:join_finetuning_tpcds}
    \end{subfigure}
    \begin{subfigure}[b]{0.48\linewidth}
        \includegraphics[width=\linewidth]{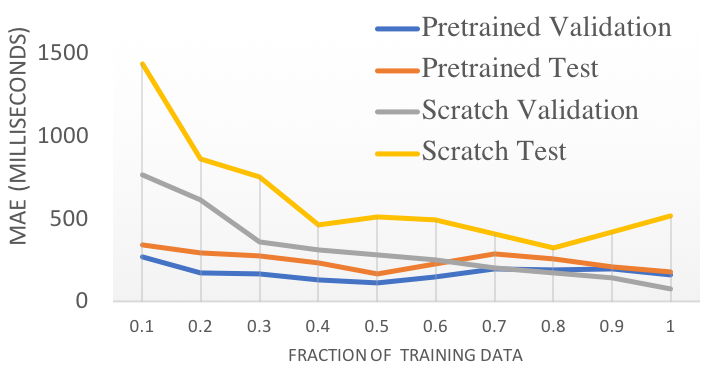}
        \vspace{-5mm}
        \caption{\small MAEs on Sort operator model for fractions of training data on TPC-DS SF-8.}
        \label{fig:sort_finetuning_tpcds}
    \end{subfigure}
    \begin{subfigure}[b]{0.48\linewidth}
        \includegraphics[width=\linewidth]{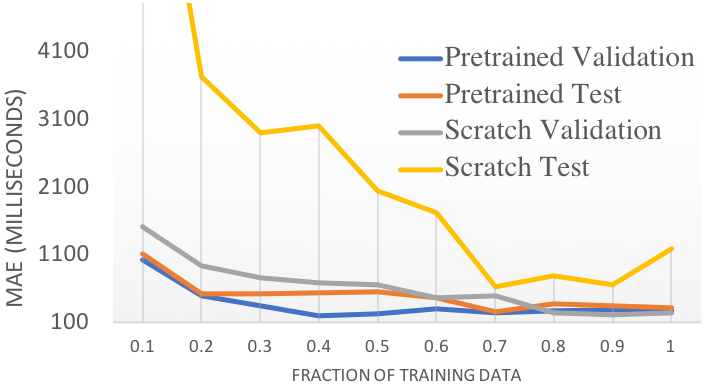}
        \vspace{-5mm}
        \caption{\small MAEs on Aggregate operator model for fractions of training data on TPC-DS SF-8.}
        \label{fig:agg_finetuning_tpcds}
    \end{subfigure}
    \vspace{-3mm}
    \caption{\small The effect of dataset size for finetuning with pretrained vs scratch(non-pretrained) models, showing $\geq$ 0.3 fraction of dataset is enough for pretrained models to adapt a new environment.}
    \label{fig:operators_finetuning}
    \vspace{-7mm}
\end{figure}

\Paragraph{Finetuning with pretrained models.} The goal of having a pretrained
model is to expedite the domain adaptability with less data. In many cases,
obtaining adequate training data is challenging and time-consuming. In this set
of experiments, we perform finetuning tasks on a new dataset, of TPC-DS with
scale factors 8 (SF-8). We also performed the same experiment on spatial
dataset which shows similar result. Due to space constraints we could not add
the result on spatial dataset.

To show the effectiveness of pretraining models over scratch or non-pretrained
model, we orchestrated a comparative experiment where the performance of models
trained on fractions of training data. We limit the full training dataset to
randomly chosen 2000 plans and test dataset to 500 plans for both TPC-DS and
Spatial datasets. We run each model for 100 epochs which takes around than 10
minutes to train.  In all the line charts from Figure
\ref{fig:operators_finetuning}, we notice that as the amount of training data
increases, the MAE decreases on all the models, but the validation MAEs of
scratch models is only comparable with the pretrained models when trained with
0.5 to 0.7 fractions of training data.  The critical observation is that
pretrained test seldom improves beyond 0.3 fractions of training data for our
workloads.

\begin{figure}[t]
    \centering
    \begin{subfigure}[b]{0.48\linewidth}
        \includegraphics[width=\linewidth]{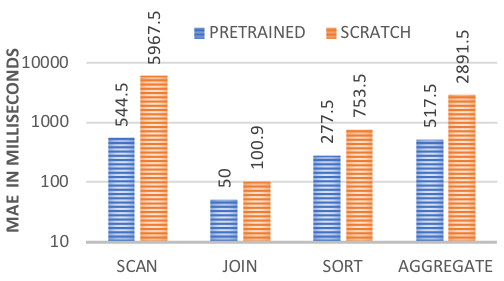}
        \vspace{-5mm}
        \caption{TPC-DS SF-8 benchmark.}
        \label{fig:pre_scratch_tpcds}
        \vspace{-2mm}
    \end{subfigure}
    \begin{subfigure}[b]{0.48\linewidth}
        \includegraphics[width=\linewidth]{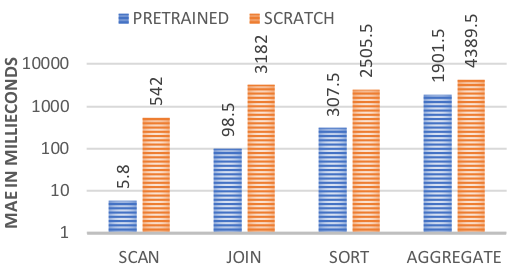}
        \vspace{-5mm}
        \caption{Spatial benchmark.}
        \label{fig:pre_scratch_spatial}
        \vspace{-2mm}
    \end{subfigure}
    \vspace{-1mm}
    \caption{Comparison of MAEs for pretrained vs scratch models with 0.3 fraction of finetuning data.}
    \label{fig:pre_vs_scratch}
    \vspace{-7mm}
\end{figure}

To make the clear distinction between pretrained and scratch models, we show
the MAE on test dataset for each operator and dataset with 0.3 fractions of
training data in Figure \ref{fig:pre_vs_scratch} for TPC-DS SF-8 and Spatial
workloads. We report the test MAE for the best validation model obtained in 100
epochs. In all the cases, the pretrained model beats the scratch model by a
considerable margin.  Conclusively, it confirms that our pretrained encoders
are useful and adapt to the new workloads quickly.

\begin{figure}[t]
    \begin{subfigure}[b]{0.48\linewidth}
        \includegraphics[width=\linewidth]{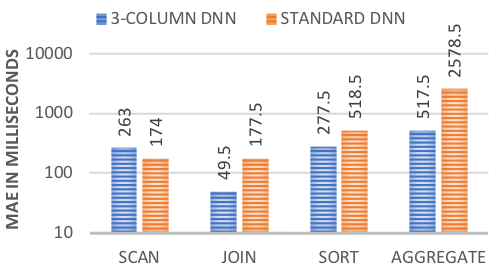}
        \vspace{-4mm}
        \caption{TPC-DS SF-8 benchmark.}
        \label{fig:model_mlp_tpcds}
        \vspace{-2mm}
    \end{subfigure}
    \begin{subfigure}[b]{0.48\linewidth}
        \includegraphics[width=\linewidth]{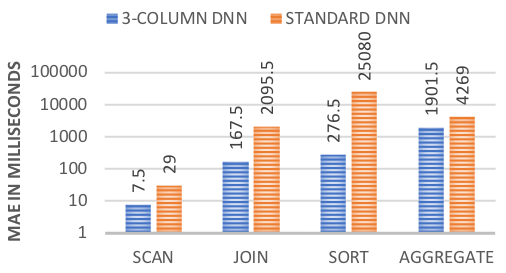}
        \vspace{-4mm}
        \caption{Spatial benchmark.}
        \label{fig:model_mlp_spatial}
        \vspace{-2mm}
    \end{subfigure}
    \vspace{-2mm}
    \caption{Comparison of MAEs for multi-column vs standard DNN models with 0.3 fraction of finetuning data.}
    \label{fig:model_vs_mlp}
    \vspace{-7mm}
\end{figure}

\Paragraph{Multi-column vs Standard DNN}
In this experiment, we perform a comparative evaluation between our
three-column DNN and a standard (single-column) DNN for the performance
encoder. Similar to the previous finetuning experiment, we pretrained both the
models with same workloads. After that, we finetuned each model with 0.3
fraction of training data from TPC-DS SF-8 and Spatial workloads independently
to obtain multiple evaluation models. Figure \ref{fig:model_mlp_tpcds} and
\ref{fig:model_mlp_spatial} shows the Mean Absolute Error(MAE) obtained from
the three-column DNN and the standard DNN models for an unseen TPC-DS SF-8 and
Spatial benchmark dataset, respectively. With the TPC-DS workload, Figure
\ref{fig:model_mlp_tpcds} shows MAE for the three-column DNN model is better
than standard DNN for all the operators except the {\em scan} operator.
Whereas, the MAE for three-column DNN is significantly less than standard DNN
for spatial workload. It suggests that keeping the performance features
($f_{node}, f_{meta}, f_{db}$) independent for the first few layers helps the
model. In the single-column standard model, different types of features might
get intertwined in the early stage of the model, impeding its learnability.

In summary, our experiments present the effectiveness of plan encoders in
learning characteristics of query plans with latency prediction task and
multiple local probes on individual encoders. The results suggest the necessity
of pretrained models to feature and understand unseen queries.  Other database
core systems certainly can leverage the plan encoders for to increase their
effectiveness and achieve {\em instance optimality}.


\vspace{-3mm}
\section{Related works}
\label{sec:related}
\Paragraph{Workload characterization.}
There exists a number of research work that uses data-driven analysis on query
plans and its features to comprehend workload characteristic
\cite{ganapathi2009predicting, akdere2012learning, ding2019ai, marcus2019plan,
li2012robust}.  Early research works \cite{ganapathi2009predicting,
li2012robust, zhu2017bestconfig}, focuses on feature engineering with data
mining techniques like k-NN\cite{cover1967nearest} on high-dimensional
features. The initial works show the importance of feature engineering, which
encourages follow up research works using neural networks for workload related
prediction tasks (metrics, resource demands, indexing, etc.)
\cite{marcus2019plan,ding2019ai,ding2018plan,li2019qtune}.

All these methods learn models from input features of query plans for a
specific task. In our paper, we show an approach to learn pretrained query plan
encoders that can be used for many downstream tasks.  Currently database
researches are proposing prepackaged AI learned models for core-components of
databases \cite{kraska2019sagedb, van2017automatic}. Our work on query plan
encoders bridges the gap between query input and prediction tasks.

Database tuning is an interesting problem to achieve instance optimality and
closely relates to query performance prediction tasks.  An earlier work, Ituned
\cite{thummala2010ituned} uses a feature-based approach for tuning database.
Recently published work, QTune \cite{li2019qtune} uses query plans and
reinforcement learning for tuning databases.  In both the approaches, query
plans are important. An attempt of ours to create a pretrained encoder for
query plans is relevant to database tuning and other similar tasks. We show its
relevancy with a latency prediction over a different configuration and
different data.  An earlier work by Popescu et al.  \cite{popescu2012same}
shows it is feasible to accomplish performance prediction tasks on new data
distribution for the same query. One of the significant contributions of our
pretrained encoders is the adaptability of the models with new query and data.



\vspace{-3mm}
\section{Conclusion}
\label{conclusion}
In this work, we study a method of featurizing database workloads with AI based
encoders that helps in understanding database queries under structural and
performance properties. We followed a pretrained encoder based approach for our
models that learns weights from diverse training dataset and then use the
learned model in downstream tasks like query latency prediction. We perfomed
multiple probes on structural encoder and performance plan encoders, to prove
their learning capability and efficacy. We also present an in-depth ablation
study on query latency prediction for multiple benchmark workload proving the
usefulness of workload characterization with plan encoders.  Our approach of
studying database workloads with pretrained encoder models paves a new
direction in this field.

\Paragraph{Acknowledgement}
\label{acknowledgement}
We would also like to show our gratitude to Hubert Lubaczewski for providing us
access to the crowdsourced plan dataset.




\bibliographystyle{abbrv}
\bibliography{main}  


%

\end{document}